\documentclass[showpacs,preprintnumbers,amsmath,amssymb,nofootinbib]{revtex4}

\usepackage{amsmath,amssymb,latexsym}
\usepackage{graphicx,bm}  
\usepackage{dcolumn}      

\begin{document}
%

\title{\bf 
Thermodynamics of four-dimensional black objects in the warped compactification
}

\author{Hideaki Kudoh } 
\email{kudoh_at_utap.phys.s.u-tokyo.ac.jp}  
\affiliation{   
   Department of Physics, The University of Tokyo, Bunkyo-ku, 
   113-0033, Japan   
} 

\author{Yasunari Kurita} 
\email{kurita_at_sci.osaka-cu.ac.jp}
\affiliation{ 
   Department of Physics, 
   Osaka City  University, Osaka 558-8585, Japan 
}

\begin{abstract}
We reinvestigate the thermodynamics of black objects (holes and strings) in four-dimensional braneworld models that are originally constructed by Emparan, Horowitz and Myers based on the anti-de Sitter (AdS) C-metric. 
After proving the uniqueness of slicing the AdS C-metric, we derive thermodynamic quantities of the black objects by means of the Euclidean formulation and find that we have no necessity of requiring any regularization to calculate their classical action. 
We show that there exist the Bekenstein-Hawking law and the thermodynamic first law.
The thermodynamic mass of the localized black hole on a flat brane is negative, and it differs from the one previously derived.   
We discuss the thermodynamic stabilities and show that the BTZ black string is more stable than the localized black holes in a canonical ensemble, except for an extreme case.
We also find a braneworld analogue of the Hawking-Page transition between the BTZ black string and thermal AdS branes.
The localized black holes on a de Sitter brane is discussed by considering Nariai instanton, comparing the study of ``black cigar" in the five-dimensional braneworld model.  
\end{abstract}


\preprint{ OCU-PHYS-214, AP-GR-16, UTAP-496 }

\maketitle

\section{Introduction}
\label{sec:intro}

In recent years, much progress has been made in investigations of black holes in higher dimensions, stimulated by proposals of various higher dimensional cosmological models. 
For the asymptotically flat vacuum case, a well-known higher dimensional black hole solution is the Myers-Perry solution that is a higher dimensional analogue of the Kerr solution in four dimensions \cite{Myers:1986un}. 
On one hand, the uniqueness of asymptotically flat and static (regular) black holes has been proved for the vacuum system in higher dimensions \cite{Gibbons:2002av}. 
The black ring solution found by Emparan and Reall is another sequence of rotating black holes \cite{Emparan:2001wn}. 
The horizon of the black ring has a different spatial section from that of the Myers-Perry solution, and the uniqueness theorem is violated for rotating black holes in higher dimensions.  
(see Refs. \cite{Kodama:2004kz,Ishibashi:2003ap} for recent studies of perturbative uniqueness of rotating black holes.)

If we take into account branes according to, e.g., the Randall-Sundrum (RS) braneworlds 
\cite{Randall:1999ee,Randall:1999vf}, little is known for black holes bound to 3-brane in five dimensions, which are expected to play a fundamental role to understand strong-gravity regime in the braneworlds. 
The exact solutions that represent such black holes are known only in the four-dimensional case: 
Emparan, Horowitz and Myers (EHM) constructed the static axisymmetric solutions by appropriately slicing the four-dimensional AdS$_4$ C-metric \cite{Emparan:1999wa,Emparan:1999fd}. 
Rotating localized black holes analogous to the Kerr black holes was also constructed by applying the AdS$_4$ D-metric \cite{Emparan:1999fd}. 
Although there is an attempt to extend the argument to higher-dimensional cases \cite{Charmousis:2003wm}, exact solutions of localized black holes have not been known so far in the original five-dimensional model. 
However, recent numerical works have succeeded in constructing such localized black holes, clarifying the physical properties of them \cite{Kudoh:2004kf}.(see also \cite{Kudoh:2003xz,Kudoh:2003vg}.)
There is an attempt constructing an analytic solution perturbatively, starting from higher-dimensional Schwarzschild black holes \cite{Karasik:2004wk}. 
However the black holes are very small, and the detailed properties of the constructed black hole are still unknown.

In this paper, we discuss the black objects (holes and strings) in four-dimensional braneworlds.
Although some parts of them have been investigated by Emparan et al. based on their solutions, they assumed the Bekenstein-Hawking law and the first law of thermodynamics in order to define the thermodynamic mass of localized black holes. 
The thermodynamics of the system in which branes and black objects coexist have not yet been well understood. 
We improve the points by adopting the traditional Euclidean formulation for thermodynamics of black holes, and we derive thermodynamic quantities from the classical action. 
As is well-known, an evaluation of a classical action often encounters infrared divergence and one needs to regularize it by some method.
In particular, one may encounter a new kind of divergence which is related to brane boundaries. 
In the present case of four-dimensional braneworlds, however, all terms that give divergences cancel out each other, and we can straightforwardly evaluate the classical action without introducing any regularization. 
We also discuss the thermodynamic stability of the black objects in a canonical ensemble which are thought to be in equilibrium with thermal radiation in the bulk. 
We show that black strings extending between two AdS branes are stable except for some extreme case in which a localized black hole is more preferable and a black string may decay into it. 
We also find the braneworld analogue of the three-dimensional Hawking-Page transition: the transition between the BTZ black string and the thermal AdS branes which are the system of the two AdS branes in thermal equilibrium with bulk radiation.
The localized black holes on a de Sitter brane are discussed by considering Nariai-instanton, comparing the study of ``black cigar" in the five-dimensional braneworld model \cite{Garriga:1999bq}.

The paper is organized as follows.
In the next section, we give a brief review of the EHM solutions in the four-dimensional braneworld models.
After we prove the uniqueness of slicing AdS$_4$ C-metric in Section \ref{sec:uniqueness of slicing}, we discuss the thermodynamics of black objects and thermal branes based on the Euclidean formulation in Section \ref{sec:Euclidaction}. 
We will find that the thermodynamic mass of the thermal AdS branes equals to that of thermal AdS$_3$ in pure 2+1 gravity.
In Section \ref{stability}, we consider the thermodynamic stabilities of black objects in a canonical ensemble, and we also discuss the Nariai instanton which is obtained from the black hole localized on the dS brane. 
We conclude with some general remarks in
Section \ref{summary}.

\section{Action and EHM solutions}
\label{sec:action and EHM sol}

The general action for $D$-dimensional vacuum solutions in the warped compactification \cite{Randall:1999ee} is given by  
\begin{eqnarray}
I_{RS} &=& I_{EH} + I_{GH} + I_{brane} ,
\label{eq:S_RS}
\\
I_{EH} &=&
 - \frac{1}{2\kappa_D} \int_{\mathcal{M} } dx^{D} \sqrt{-g} \left( R - 2\Lambda \right)  ,
\cr
I_{GH} &=& - \frac{1}{\kappa_D} \int_{\mathcal{\partial M} } dx^{D-1} \sqrt{-h} K ,
\cr
I_{brane}  &=&  \sum_{k=1, 2, \cdots } \sigma_k \int_{\mathcal{\partial M} } dx^{D-1} \sqrt{-h},
\nonumber
\end{eqnarray}
where  $\Lambda_D := - (D-1)(D-2)/2\ell^2_D$ is the bulk cosmological constant. 
$\kappa_D = 8\pi G_D$ is $D$-dimensional gravitational constant.
The first term in $I_{RS}$ is the usual Einstein-Hilbert action with a negative cosmological constant. 
The second term is the Gibbons-Hawking boundary term, which is necessary for a well-defined variation problem. 
$K$ is the trace of the extrinsic curvature of the boundary; 
$K_{\mu\nu}=h^\rho_\mu h^\sigma_\nu \nabla_\rho n_\sigma$, 
where $h^\nu_\mu$ is defined as $h^\nu_\mu = \delta^\nu_\mu - n_\mu n^\nu$, 
and $n^\nu$ denotes an ``outward" unit normal to the boundary
\footnote{
In the brane-world context, ``inward'' unit normal vector is commonly used to derive Israel's junction condition at the brane.  
In this paper, we adopt the standard outward unit notation (see e.g. Ref. \cite{Chamblin:1999ya,Hawking:2000kj} )
}. 
The $I_{brane}$ is the action of branes with tensions $\sigma_k$.  
As for the RS construction, the tensions are given by $ \sigma_\pm := \pm 2(D-2)/\kappa_D \ell_D$ for the branes with $\pm$-tension.

$Z_2$-symmetry is imposed at each brane in the warped compactification. 
Boundary conditions at those surfaces are given by Israel's junction condition. 
The surface terms in the action (\ref{eq:S_RS}) gives the junction condition 
\begin{eqnarray}
  [K_{\mu\nu} - K h_{\mu\nu}]
 =  \kappa_D T_{\mu\nu},  
\label{eq:Israel conditions 1}
\end{eqnarray}
where $T_{\mu\nu} = - \sigma h_{\mu\nu}$, and $[F(z)]$ denotes the difference of $F$ on the boundary at $z$: $[F]=\lim_{\epsilon \to 0} F|_{z+\epsilon} - F|_{z-\epsilon}$.
Thus the boundary condition that any vacuum solutions must satisfy is 
\begin{eqnarray}
K_{\mu\nu}|_{\epsilon+} - \frac{\sigma \kappa_D }{2(D-2)} h_{\mu\nu} =0, 
\label{eq:Israel conditions 2}
\end{eqnarray}
where we have used $Z_2$-symmetry. From this result we notice that the Gibbons-Hawking term evaluated on the brane is related to the brane action by 
\begin{eqnarray}
 I_{brane} = - \frac{ D-2 }{ D-1 } ~  I_{GH}|_{brane}. 
 \label{eq:I_brane and I_GH}
\end{eqnarray}

As was shown in \cite{Emparan:1999fd}, one can construct a black hole bound to a brane in the four-dimensional RS model from AdS C-metric.
The general AdS$_4$ C-metric is 
\begin{eqnarray}
 ds^2 = \frac{1}{A^2 (x-y)^2}
  \left[
   - F(y) dt^2 +\frac{1}{F(y)} dy^2 + \frac{1}{G(x)}dx^2 + G(x) d\phi^2
  \right],
  \label{eq:AdS Cmetric}
\end{eqnarray}
where 
\begin{eqnarray*}
 G(x) &=& 1 + k x^2 - 2 m A x^3 ,
\cr
 F(y) &=& 
  \lambda - ky^2 +2m A y^3 .
\end{eqnarray*}
The parameters $m$, $A$ and $k=0,\pm 1$ are related to the mass, acceleration and topology of the black hole, respectively.
The metric satisfies 
\begin{eqnarray}
R_{\mu\nu} &=& - \frac{3}{\ell_4 ^2 } g_{\mu\nu} ,
\cr
\ell_4 &=& \frac{1}{A \sqrt{\lambda + 1} },
\label{eq:ell4}
\end{eqnarray}
and then $\lambda$ is thought to be related to the bulk cosmological constant.
Note that from Eq.(\ref{eq:ell4}), we require $\lambda \geq -1$.
Scalar curvatures are given by
 \begin{eqnarray}
 R^{\mu\nu\rho\lambda} R_{\mu\nu\rho\lambda}
 &=& 24 A^4 [ 2A^2 m^2(x-y)^6+(1+\lambda)^2 ] \,,
 \cr
 C^{\mu\nu\rho\lambda} C_{\mu\nu\rho\lambda} 
 &=& 48 A^6 m^2 (x-y)^2
  \,.
 \label{eq:curvature}
\end{eqnarray}
Hence a curvature singularity does not appear as long as the parameters and the coordinate values are finite.

\subsection{Black holes in two branes system (EHM II) }
\label{subsec:EHM II}

In this subsection we review axially symmetric static black holes bound to an asymptotically AdS$_3$ brane in the system of two branes. 
We call such black holes the EHM II solutions, whereas the black hole bound to an asymptotically flat brane in the system of single brane is called the EHM I solution.
The reader who is familiar to these black hole solutions can skip straight to the next section.

The branes can be placed at $x=0$ and $y=0$ surfaces in the AdS C-metric (\ref{eq:AdS Cmetric}), 
where the junction conditions (\ref{eq:Israel conditions 2}) are satisfied.
The cosmological constants on the branes are 
\begin{eqnarray}
\Lambda _3 = - \frac{\lambda}{L^2} \,,
\label{eq:cosmological const on the brane}
\end{eqnarray}
where $L=L_1 := 1/A$ at $x=0$ and $L=L_2 := \sqrt{\lambda} /A$ at $y=0$. Here we assume that $\lambda$ is positive. 
The slicing implies the following tensions of the branes, 
\begin{eqnarray}
\kappa_4 \sigma|_{x=0} &=& 4A \sqrt{G(0)} = 4A ,
\cr
\kappa_4 \sigma|_{y=0} &=& 4A \sqrt{F(0)} = 4A \sqrt{\lambda}.
\label{eq:tension}
\end{eqnarray}
In the present case, we also find
\begin{eqnarray}
G_{3} &=& \frac{A G_4}{2}, 
\\
\ell_3^2 &=& \frac{1}{A^2 \lambda}, 
\label{eq:ell_3}
\end{eqnarray}
where $G_3$ is the effective gravitational constant and $\ell_3$ is the AdS radius. 
Both of them are normalized on the brane at $x=0$.
On the axis of rotation,  $G(x)$ vanishes and the equation has one positive root $x_2$ for $m>0$. 
To avoid a conical singularity on the axis $x=x_2$, the period of $\phi$ must be 
\begin{eqnarray}
 \Delta \phi = \frac{4\pi}{ |G'(x_2)|}.
 \label{eq:deficit angle}
\end{eqnarray}
In this construction of a localized black hole, $x$ is restricted to lie in the range $0 \le x \le x_2$.  
Because of the overall factor $(x-y)^{-2}$ in (\ref{eq:AdS Cmetric}), $y$ is restricted to satisfy $-\infty \le y \le x$.  
The smallest zero of $F(y)$, $y_0$, defines the black hole event horizon. 
From Eq. (\ref{eq:curvature}) it turns out that a curvature singularity is only at $y= - \infty$ which is enclosed by the event horizon.

By taking $\rho=-1/y$, the geometry on the brane at $x=0$ is 
\begin{eqnarray}
 ds^2 = \frac{1}{A^2}
 \left[
 - \left( \lambda \rho^2 - k - \frac{2m A}{\rho} \right) dt^2
 + \left( \lambda \rho^2 - k - \frac{2m A}{\rho} \right)^{-1} d\rho^2
 + \rho^2 d\phi^2 
 \right].
\label{eq:metric on the x=0 EHMII}
\end{eqnarray}
This is similar to the nonrotational BTZ black hole, except the extra terms of $2m A /\rho$ and the periodicity of $\phi$.
To compare the induced metric with the BTZ metric, we introduce the following normalized coordinates so that it has asymptotically AdS$_3$ form with the proper angular periodicity $2\pi$,
\begin{eqnarray}
\hat{t}    = \frac{2\pi}{A  \Delta \phi} t,
~~
\hat{\rho } = \frac{ \Delta \phi}{2\pi A}  \rho,
~~
\hat{\phi} = \frac{2\pi}{ \Delta \phi} \phi.
\label{eq:rescaled coord}
\end{eqnarray}
Then the induced metric (\ref{eq:metric on the x=0 EHMII}) becomes
\begin{eqnarray}
 ds^2 = 
 - \left[   \lambda A^2 \hat{\rho}^2 - k \left( \frac{\Delta \phi}{2\pi} \right)^2
          - \frac{2m }{\hat{\rho}} \left( \frac{\Delta \phi}{2\pi} \right)^3
 \right] d\hat{t}^2
 + \left[   \lambda A^2 \hat{\rho}^2 - k \left( \frac{\Delta \phi}{2\pi} \right)^2
          - \frac{2m }{\hat{\rho}} \left( \frac{\Delta \phi}{2\pi} \right)^3
 \right]^{-1} d\hat{\rho}^2
 + \hat{\rho}^2 d\hat{\phi}^2 ,
\label{eq:metric on the x=0 case2}
\end{eqnarray}
and the periodicity of $\hat{\phi}$ is $2\pi$.

The AdS$_4$ C-metric with the above slicing can represent other solutions by setting $m=0$ 
and/or permitting negative $\lambda$. 
In the latter case, we cannot consider the second brane due to Eq. (\ref{eq:tension}). 
When $m=0$, $\lambda >0$ and $k=+1$, the metric shows a BTZ black string (BS) that extends between two AdS branes. 
In the case of $m=0, k=-1$, the induced metric (\ref{eq:metric on the x=0 EHMII}) is not a black hole solution, but a solution of three-dimensional Einstein's equations with a cosmological constant;
\begin{eqnarray*}
    {}^{(3)}G^\mu_\nu = A^2 \lambda g^\mu_\nu. \quad (m=0, k=-1)
\end{eqnarray*}
Then for $\lambda>0$, the brane is AdS brane, whereas it is dS brane for $\lambda<0$.
These solutions serve as background spacetimes containing black holes.
A classification of the black hole solutions is summarized in Table \ref{table:Classification}.

\begin{table}[h]
\begin{center}
\begin{tabular}{r cc c|lll}
\hline   \hline
   & $m$ & $\lambda$  & $k$ & 
\\   \hline
   & $m=0$ & $\lambda>0$ & $k=+1$ &   BTZ BS and AdS branes
\\ & & $\lambda>0 ~(<0)$ & $k=-1$& AdS (dS) branes  
\\ & & $\lambda=0 $ & $k=-1$&a flat  brane \\
\hline
 & $m>0$  & $\lambda>0$ & $k=0, \pm 1$ & BH and two AdS branes (EHM II)\\ 
 & &$-1<\lambda<0$ & $k=-1$& BH on a dS brane \\
 & &$\lambda=0$ & $k=-1$& BH on a flat brane (EHM I) \\
\hline \hline   
\end{tabular}
\caption[short]{
Classification of black objects.
}
\label{table:Classification}
\end{center}
\end{table}

\subsection{Black hole localized on a flat brane  (EHM I)}

A black hole bound to a flat brane in the system of a single brane was discussed in Ref. \cite{Emparan:1999wa} as the simplest slicing of AdS C-metric.  
The black hole can be formally obtained by taking $\lambda \to 0$ in the construction of localized black holes in the system of two branes. 
In the limit, the brane at $y=0$ formally vanishes because the tension vanishes. 
The surface $y=0$ then corresponds to AdS (Cauchy) horizon in the one-brane system. 
The point $y=x$ is infinitely far away from points with $y \neq x$.

A black hole solution is possible for $k= -1$. The induced metric on the flat brane at $x=0$ is 
\begin{eqnarray}
 ds^2 =  
 - \left( 1 - \frac{2m }{r } \right) d\tilde{t}^2
 + \left( 1 - \frac{2m }{r } \right)^{-1} dr^2
 + r^2 d\phi^2 ,
\label{eq:metric on the x=0 EHMI}
\end{eqnarray}
where $r= \rho /A$ and $\tilde{ t }= t/A$.
This is similar to the Schwarzschild black holes, but is not a vacuum solution in pure 2+1 gravity. 
The periodicity of $\phi$ is given by Eq. (\ref{eq:deficit angle}).

\section{Uniqueness of slicing}
\label{sec:uniqueness of slicing}

In contrast to the situation of no black hole on a brane, 
the condition of placing a brane becomes very restrictive in the presence of a black hole \cite{Chamblin:1999by,Kodama:2002kj}.
Although we have discussed possible constructions of localized black holes in the previous section, 
one might think that other slicing could be allowed to construct unknown black hole solutions. 
A possible case is discussed in Ref. \cite{Emparan:1999fd}: any surface $x=x_b$ with $G'(x_b) =0$ and $G(x_b) \neq 0$, or 
$y=y_b$ with $F'(y_b) =0$ and $F(y_b) \neq 0$, are allowed as a consistent slicing of AdS C-metric. 
However it is shown that such slicings are identical to the one of $x=0$ and $y=0$ discussed in the previous section, after shifting coordinates and redefining parameters \cite{Emparan:1999fd}.
In this section, we prove that there is no other possible slicing, except a physically uninteresting case.

We consider a general situation in which the location of a brane is given by $y=y(x)$ 
in the spacetime (\ref{eq:AdS Cmetric}) \cite{Chamblin:1999by}. 
The timelike tangent to the brane is 
\begin{eqnarray}
    u = \frac{A(x-y)}{\sqrt{F}} \frac{ \partial}{\partial t}, 
\end{eqnarray}
and the spacelike tangents are 
\begin{eqnarray}
v &=& \frac{1}{C} 
\left(   y'\frac{ \partial}{\partial y} + \frac{\partial}{\partial x}  
\right),
\cr
e_{\phi} &=&
 \frac{A(x-y)}{\sqrt{G}}  \frac{ \partial}{\partial \phi}, 
\end{eqnarray}
where 
$C = 
 \frac{1}{A(x - y)}\sqrt{ G^{-1} + {y'}^2/{F}}
$.
Then the unit normal to the brane can be written as
\begin{eqnarray}
n
= 
\frac{ s A(x-y)}{ \sqrt{F+ (y')^2 G}}
\left( F\frac{\partial}{\partial y}-y' G\frac{\partial}{\partial x} \right),
\end{eqnarray}
where $s=\pm 1$ is the sign of the normal vector.

Let us consider the following components of the extrinsic curvature on the brane,
\begin{eqnarray}
K_{tt} &=& - \frac{ 
    s F \left[2F +(x-y)F' + 2 y'G \right]
}{2A(x-y)^2 \sqrt{F+G (y')^2}}, 
\cr
K_{\phi \phi} &=& \frac{ 
   s G \left[ 2F - (x-y)G'y' + 2y'G \right]
}{2A(x-y)^2 \sqrt{F+G (y')^2}}. 
\end{eqnarray}
From these components and the boundary conditions (\ref{eq:Israel conditions 2}), we obtain an equation by eliminating a tension of the brane, 
\begin{eqnarray}
s (x-y) (F'(y)+G'(x)y') =0.
\end{eqnarray}
Solutions of this equation are $y=0$, $y=x$ and 
$y(x)= c k x/\bigl( k +3 (c-1)A  m x  \bigr)$ ($c \neq 0,~1  $). 
If the integration constant is $c=1$, then the third solution is identical to the second one.

Now we examine if these solutions satisfy the junction condition (\ref{eq:Israel conditions 2}) with an appropriate tension. 
For the solution $y=0$, we find that a slicing is possible for $\sigma = 4s A \sqrt{\lambda}/\kappa_4$. 
This is the slicing used in the EHM II solution. 
One can also show that the junction conditions are satisfied for the solution $y=x$ with a tension $\sigma = 4s A \sqrt{1+\lambda}/\kappa_4$. 
However the brane on this surface is infinitely far away from any points in the bulk ($y\neq x$), and then this slicing is physically uninteresting. 
The remaining possibility is the third solution. It is verified that the third solution is not consistent with the junction condition.
To see the inconsistency, it is enough to study the junction conditions around $x\approx 0$. Expanding the junction conditions near the $x=0$, we find that the left hand side of (\ref{eq:Israel conditions 2}) diverges at $x=0$ and the equation cannot be satisfied for any choice of tension. 
After all, except for $y=0$ and $x=0$ slicing, there is no other physically interesting slicing of AdS$_4$ C-metric that allows localized black holes.

\section{Euclidean action and thermodynamics }
\label{sec:Euclidaction}

We now turn to discuss the thermodynamics of the black objects in the Euclidean formulation, which is a traditional formulation to yield thermodynamic quantities of black holes from their classical actions \cite{Gibbons:1977ue}.  
In computation of classical actions, one usually encounters infrared divergences and then some suitable regularization is required.
As for the traditional method, we need to subtract an appropriate background solution.  
On the other hand, a so-called counterterm method has been developed in the case of asymptotically AdS spacetime 
 \cite{Henningson:1998ey,Balasubramanian:1999re,Emparan:1999pm}. 
At first sight, some regularization seems to be required for the black objects that we are considering. 
However, as we will see below, the infrared divergences appearing from each term of the total action (\ref{eq:S_RS}) cancel out, and surprisingly, there is no need to subtract any background or introduce any counterterms.

\subsection{EHM II ($\lambda >0, m>0$) }

\subsubsection{ Euclidean action }

In order to evaluate the classical action, we need to fix the periodicity of the Euclidean time $ \tau=it$ which is given by requiring the regularity at the event horizon or the origin of the Euclidean section.
Then we get the period $\Delta \tau$ as
\begin{eqnarray}
  \Delta \tau = \frac{4\pi}{|F'(y_0)|}  .
\label{eq: d tau for EHMII BH}
\end{eqnarray}
Taking into account the $Z_2$-symmetry and the junction condition, the action for the black objects discussed in Sec. \ref{subsec:EHM II} is reduced to 
\begin{eqnarray}
I_{EH} 
&=& \frac{6}{\kappa_4 \ell^2} \int^{\Delta \tau}_{0} d\tau \int^{\Delta\phi}_{0} d\phi 
    \int^{x_2}_{ \epsilon_x } dx
    \int^{ -\epsilon_y }_{y_0} dy \sqrt{g} , 
\cr
&=&
 \frac{\Delta \tau ~\Delta \phi }{\kappa_4 \ell^2 A^4 }
 \left[
- \frac{1}{(x_2+\epsilon_y)^2} + \frac{1}{(x_2 - y_0)^2}
+ \frac{1}{(\epsilon_x + \epsilon_y)^2}
- \frac{1}{(\epsilon_x - y_0)^2}
\right] , 
\cr
I_{brane} + I_{GH}|_{(brane)}&= &
\frac{\Delta \tau ~\Delta \phi }{\kappa_4  A^2}
\left[
  \frac{F(y)}{(x_2-y)^2} - \frac{F(y)}{(\epsilon_x-y)^2}
\right]_{y=-\epsilon_y}
+
\frac{\Delta \tau ~\Delta \phi }{\kappa_4  A^2}
\left[
  \frac{G(x)}{(x-y_0)^2} - \frac{G(x)}{(x+\epsilon_y)^2}
\right]_{x=\epsilon_x} ,
\end{eqnarray}
where we have used Eq. (\ref{eq:I_brane and I_GH}) and introduced temporal cutoff parameters $\epsilon_x$ and $\epsilon_y$.
Then the total action is
\begin{eqnarray}
I_{RS}
=
\frac{\Delta \tau ~\Delta \phi }{\kappa_4  A^2 }
\left(
      \frac{1 + \lambda}{(x_2 - y_0)^2}
    - \frac{\lambda }{y_0^2} - \frac{1}{x_2^2}  
\right)  ,
\label{eq:I_EHM2_cal2}
\end{eqnarray}
where we have used (\ref{eq:ell4}) and have taken the limit $\epsilon_x, \epsilon_y \to 0$.
The period of the angular coordinate $\Delta\phi$ is given by Eq. (\ref{eq:deficit angle}).  
As we notice, the total action is finite in the limit. All divergences which come from $I_{EH}$, $I_{brane}$ and $I_{GH}$ cancel out.

It is convenient to introduce an auxiliary variable as in Ref. \cite{Emparan:1999fd}
\begin{eqnarray}
 z  = \frac{|y_0|}{x_2}, 
\end{eqnarray}
for which one has various identities such as 
\begin{eqnarray}
x_2  &=& \frac{1}{z} \sqrt{ \left| \frac{\lambda - z^3}{k(1 + z)} \right|}, 
\cr
|y_0| &=&  \sqrt{ \left| \frac{\lambda - z^3}{k(1 +z)} \right|} ,
\cr
2m A &=&  {z (z^2+\lambda) \sqrt{ 1 +z }}  \left| \frac{k}{\lambda - z^3} \right| ^{3/2} ,
\end{eqnarray}
where $k \neq 1$. 
Substituting these into the total action, we finally obtain 
\begin{eqnarray}
I = - \frac{8\pi^2}{\kappa_4 A^2} z 
\left( 
    \frac{1}{3z^2+2z^3 +\lambda} + \frac{z}{z^3 +2 \lambda +3\lambda z }
\right) .
\label{eq:Euclidean action result}
\end{eqnarray}
Here $0\le   z < \lambda^{1/3}$ for $k=+1$, 
and $\lambda^{1/3} <  z \le \infty$ for $k=-1$. 
In the case of $k=0$, we think that 
$ z= \lambda^{1/3}$ because $x_2 = 1/(2mA)^{1/3}$ and $|y_0| = (|\lambda|/2mA)^{1/3} $ for $\lambda>0$.

\subsubsection{ Thermodynamic quantities }

\begin{figure}
\begin{center}
   \includegraphics[width=7cm,clip]{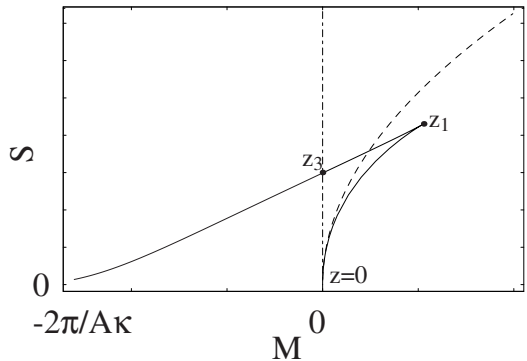}
\end{center}
\caption{ 
\label{fig:SvsM_lambda3}
Schematic figure of the relation between the entropy $S$ and the thermodynamic mass $M$ for the BTZ black string (dotted line) and the black hole (solid line).  
The global feature of the relation does not depend on $\lambda >0$. 
}
\end{figure}

Let us consider the canonical ensemble of the black objects, i.e., the situation in which the system is in equilibrium with thermal radiation at the Hawking temperature of the black objects. The free energy is defined by 
\begin{eqnarray}
  F &=& \frac{I}{\beta},
\end{eqnarray}
where $\beta$ is related to the Hawking temperature $T$ as $T=\beta^{-1}$.  
Then the thermodynamic mass $M$ of the system and the entropy are derived from the classical action through the free energy, 
\begin{eqnarray}
  M =  \partial_{\beta} I,
\quad 
  S = \beta  M   - I.
\label{eq:M and S}
\end{eqnarray}
The Hawking temperature is given by the inverse of the period of the Euclidean time. We adopt 
Eq. (\ref{eq:rescaled coord}) as the normalization of the Killing time, thus
\begin{eqnarray}
  \beta = \frac{2\pi}{A \Delta \phi} \Delta \tau. 
\end{eqnarray}
In terms of the auxiliary variable $z$, thermodynamic quantities are expressed as  
\begin{eqnarray}
\beta 
 &=& \frac{2\pi}{A z} \frac{\lambda +3 z^2 +2 z^3}{2\lambda +3 \lambda z + z^3} , 
\cr
  F &=&   - \frac{8\pi z^2 (\lambda +2 z \lambda +2z^3 + z^4 )}
           {A\kappa_4 (\lambda + 3z^2 + 2z^3)^2},
\cr
   M  &=& \frac{8\pi z^2(1+z)(\lambda -z^3)}{A \kappa_4 (\lambda +3z^2 +2z^3)^2 } ,
\cr
   S &=&   \frac{16\pi^2 z}{A^2 \kappa_4 (\lambda +3z^2 +2z^3)} .
\label{thermalQ:EHM2}
\end{eqnarray}
One can verify that the entropy is identical to one quarter of the horizon area ${\mathcal {A} }$
in the Planck unit, 
\begin{eqnarray*}
   S = \frac{ {\mathcal {A} }}{4G_4},
\end{eqnarray*}
recovering the well-known Bekenstein-Hawking law. 
We can easily confirm the first law of thermodynamics for this system.
It also leads to the fact that the thermodynamic mass $M$ is identical to the (four-dimensional) mass in Ref. \cite{Emparan:1999fd} obtained by integrating the first law. 
From Eq. (\ref{thermalQ:EHM2}), we notice that the thermodynamic mass of the black hole for $z<\lambda^{1/3} $ $(k=+1)$ is positive, whereas it is negative for $z>\lambda^{1/3}$ $(k=-1)$ (see Fig. \ref{fig:SvsM_lambda3}).  
The minimum mass of the EHM II black hole is 
\begin{eqnarray}
    M_{min}=-\frac{2\pi}{\kappa_4A} = -\frac{1}{8G_3}, 
\end{eqnarray}
which corresponds to the limit:
$z \to \infty $ ( or $m\to 0$).
In this limit, the solution represents two AdS branes without black objects.

A special case is $k=0$ and $\lambda>0$.  
We can think that this solution connects the solutions for $k=1$ with those for $k=-1$ \cite{Emparan:1999fd}.
Indeed, thermodynamic quantities of this special solution tie those of them; 
\begin{eqnarray}
\beta &=& \frac{2\pi}{A \lambda^{2/3}} , 
\cr
F &=&  - \frac{8\pi  \lambda^{1/3} }
           {3A\kappa_4 (1+ \lambda^{1/3} )},\quad  
\cr
M &=&  0 ,
\cr
S &=&  \frac{16 \pi^2}{3 \kappa_4 A^2} \frac{1 }{\lambda^{1/3} (1+ \lambda^{1/3} )} .  
\end{eqnarray}
The thermodynamic mass equals zero, and it corresponds to the point $z_3$ in Fig. \ref{fig:SvsM_lambda3}.

We can consider AdS branes with finite temperature, i.e., the AdS branes that are in thermal equilibrium with thermal radiation in the bulk. We call this system ``the thermal AdS branes." The classical action of the thermal AdS branes is 
\begin{eqnarray}
I=-\frac{2\pi \Delta\tau}{\kappa_4 A^2},
\label{AdSbrane-action}
\end{eqnarray}
where the period $\Delta \tau$ of the Euclidean time is related to the temperature of thermal radiation 
as follows:
\begin{eqnarray}
\beta_{brane}=\frac{\Delta\tau}{A}.
\label{branebeta}
\end{eqnarray}
Here $\beta_{brane}$ is the inverse temperature of the thermal radiation. 
Then the thermodynamic quantities of the thermal AdS branes are
\begin{eqnarray}
F_{brane} &=& 
-\frac{1}{8G_3}, \cr
M_{brane} &=& 
-\frac{1}{8G_3},\cr
S_{brane} &=& 0 \label{braneS}.
\end{eqnarray}
These are constant for any temperature.
We should note that the negative mass $M_{brane}$ equals that of AdS$_3$ which corresponds to the Neveu-Schwarz (NS) ground state of the dual conformal field theories (CFT) and can be interpreted as the Casimir energy of the CFT 
\cite{Coussaert:1994jp,Balasubramanian:1999re}. 
Because the thermodynamic mass of the thermal AdS branes is determined without subtracting any background, it may have some definite meanings as in the case of AdS$_3$. 
This fact implies that there is some natural interpretation of the negative ground mass 
in terms of cutoff CFT corresponding to the four-dimensional brane system.

As discussed in Ref. \cite{Emparan:1999fd}, the negative mass sector $(k=-1)$ of these localized black holes interpolates between the AdS brane and the massless black hole $(k=0)$.
In the pure 2+1 gravity with negative cosmological constant, these solutions have naked conical singularities \cite{Banados:1992wn}.
But, in the four-dimensional braneworld, the conical solution changes into a black hole with regular horizon. 
It might be an outcome of the backreaction of quantum effects of the corresponding CFT as was discussed in Ref. \cite{Emparan:2002px, Tanaka:2002rb}.

\subsection{BTZ black string $(m=0, k=+1, \lambda>0)$}

The BTZ black string is the solution with $m=0$.  
The classical action for the solution can be obtained from (\ref{eq:I_EHM2_cal2}) 
by taking $y_0=\sqrt{\lambda}$ and $x_2 \to \infty$. The result is  
\begin{eqnarray*}
 I = - \frac{2\pi \Delta \phi_{BTZ} }{\kappa_4 A^2 \sqrt{\lambda}} ,
\end{eqnarray*}
where we have used $\Delta \tau = 2\pi/\sqrt{\lambda}$.  
Because there is no axis in this case, any period $\Delta \phi_{BTZ}$ is possible.  
The period is the parameter specifying the BTZ black string and related to the three-dimensional mass as,
\begin{eqnarray}
    M_3 = \frac{1}{8G_3}\left(\frac{\Delta\phi}{2\pi}\right)^2.
\end{eqnarray} 
Thermodynamic quantities derived from the action are 
\begin{eqnarray}
 {\beta}_{BTZ}
 &=& \frac{4\pi^2}{A \Delta\phi_{BTZ} \sqrt{\lambda} } ,
\cr
  F_{BTZ} &=& - \frac{\Delta\phi_{BTZ}^2}{2\pi A \kappa_4 } ,
\cr
  M_{BTZ}  
  &=& \frac{\Delta\phi_{BTZ}^2}{2\pi \kappa_4 A } ,
\cr
    S_{BTZ}
      &=& \frac{4 \pi \Delta \phi_{BTZ}}{ A^2 \kappa_4 \sqrt{\lambda}} 
    = \frac{ {\mathcal{ A }} }{4G_4},
\end{eqnarray}
where ${\mathcal{A}}$ is the total area of the horizon, recovering again the Bekenstein-Hawking law. 
The thermodynamic mass and entropy are identical to those obtained in Ref. \cite{Emparan:1999fd}.

\subsection{EHM I ($\lambda =0, k=-1$)}

In the case of the single brane model, there is a Cauchy horizon in the bulk.
To evaluate the Euclidean action, we only consider a domain within the Cauchy horizon. 
The calculation of the Euclidean action is the same as before. 
The contribution from the Gibbons-Hawking term on the Cauchy horizon vanishes,
\begin{eqnarray}
I_{GH}|_{y=0} &=& 0.
\end{eqnarray}
Then resulting action can be formally obtained from Eq. (\ref{eq:Euclidean action result}) 
by taking $\lambda \to 0$ $(k=-1)$ 
\begin{eqnarray} 
I &=& 
 - \frac{ 16\pi^2 }{\kappa_4 A^2} 
   \frac{ z + 2 }{ z (2 z + 3 )}. 
\end{eqnarray}  
The associated thermodynamic quantities are 
\begin{eqnarray}
\beta &=& \frac{2\pi }{A z}\left( 2+ \frac{3}{z} \right), 
\cr  
 F &=& - \frac{ 8\pi }{A\kappa_4} 
   \frac{z (z+2)}{(2z + 3 )^2},
\cr  
 M &=& - \frac{8 \pi  (1+ z) z}{A \kappa_4  (2z + 3)^2},
\cr  
 S &=& \frac{16 \pi^2 }{A^2 \kappa_4} \frac{1}{z (2 z + 3)} 
 = \frac{ {\mathcal{A}}}{4 G_4}.   
\label{eq:EHM1-M, S}
\end{eqnarray}
As in the case of the EHM II, we can check that there exists the first law of thermodynamics for the EHM I black hole as well as the Bekenstein-Hawking law.

We notice that the thermodynamic mass (\ref{eq:EHM1-M, S}) differs from the one obtained in Ref. \cite{Emparan:1999wa}.  
The derived thermodynamic mass is negative. Moreover, not only mass but also thermodynamic properties are different from the original ones. 
The reason for these differences is that we have adopted the different normalization of timelike Killing vector
\footnote{
In Ref. \cite{Emparan:1999wa}, the "three-dimensional" mass is defined by the formula in pure 2+1 gravity, in which a mass is determined by a deficit angle. 
It was shown that the three-dimensional mass is identical to the "four-dimensional" mass derived by assuming the first law. 
However, the definition of the Hawking temperature in Ref. \cite{Emparan:1999wa} is inconsistent with Ref. \cite{Emparan:1999fd} because of the different normalization of the timelike Killing vector.
(Our normalization is consistent with that in Ref. \cite{Emparan:1999fd}.)
Hence the physical meaning of the four-dimensional mass is ambiguous and the reason for the agreement is not clear. 
Nevertheless, the agreement is surprising.
}.
We have normalized the coordinate as in (\ref{eq:rescaled coord}) so that the angular coordinate $\hat{\phi}$ has the period $2\pi$. 
If we use original (Euclidean) time coordinate $\tau$ to define the temperature as in Ref. \cite{Emparan:1999wa}, we cannot reproduce the Bekenstein-Hawking law, which means that the normalized coordinates are preferable to define the Hawking temperature. 
Moreover we can easily show that there is no other choice of temperature that produces the Bekenstein-Hawking law, except for a degree of rescaling.

The EHM I solution can be thought of as a remnant of the negative mass sector of EHM II in the $\lambda \to 0$ $(z_3 \to 0)$ limit. 
As in the case of EHM II, the minimum mass solution is the flat brane without black holes. 
The thermodynamic mass of the system is 
\begin{eqnarray}
M_{min}=-\frac{2\pi}{\kappa_4A} = -\frac{1}{8G_3}.
\end{eqnarray}
Although there is no cosmological constant on the brane, the mass of this system coincides with that of AdS$_3$ which does not depend on the AdS radius or cosmological constant but does just on the three-dimensional Newton constant $G_3$.
The agreement is surprising because the single brane system is locally AdS$_4$.

We can also consider a flat brane with finite temperature as in the case of EHM II.
The classical action of the thermal flat brane is identical to that of AdS brane (\ref{AdSbrane-action}).
Then the thermodynamic quantities of the thermal system of the flat brane are also equal to those of the thermal AdS branes
(\ref{branebeta}) and (\ref{braneS}).
Thus, the negative mass of the EHM I black hole is bounded below by that of the flat brane.

\section{Thermodynamic stabilities of black objects}
\label{stability}

\subsection{Comparison between the localized black hole and the  BTZ black string in a canonical ensemble}

\begin{figure}
\begin{center}
  \includegraphics[width=7cm,clip]{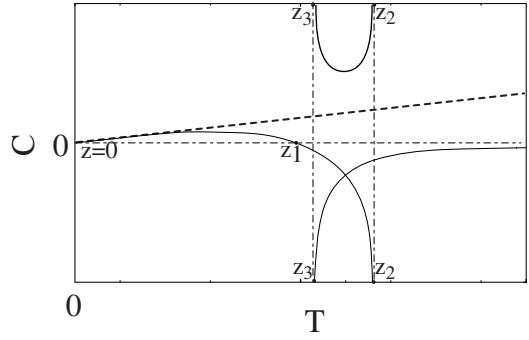}
\end{center}
\caption{ 
\label{fig:CvsT}
Schematic figure of the relation between temperature $T$ and specific heat $C$.
The BTZ black strings and the black holes for $0<\lambda<1$ are shown by dashed line and solid line, respectively.
In the case of $\lambda>1$, the $ z_2$ exchanges the position with that of $ z_3$ without changing other features. 
One can see that the specific heat is positive in the two regions, $0<z<z_1$ and $z_2<z<z_3$.
}
\end{figure}

For an isolated system of a black object, the stability of a thermodynamic system is studied by analyzing the behavior of the entropy (Fig.~\ref{fig:SvsM_lambda3}).
In this section, we investigate thermodynamic stability of black objects in a canonical ensemble, i.e., in thermal equilibrium with thermal radiation in the bulk.  
For that purpose, we compare the free energy of the localized black hole with that of the BTZ black string.
The specific heat of the BTZ black string is positive, which is given by 
\begin{eqnarray}
 C_{BTZ} = T \frac{\partial S}{ \partial T} = \frac{16 \pi^3 T}{A^3 \kappa_4 \lambda}.
\end{eqnarray}
Thus, the BTZ black string can be in stable equilibrium.
On the other hand, however, the specific heat of the localized black holes is not always positive.
It is given by 
\begin{eqnarray}
 C_{BH} = \frac{8\pi^2 z [ z^2(3+4z) - \lambda] [z^3 +2\lambda +3z \lambda]}
{A^2 \kappa_4 (\lambda - z^3) [ z^2 (3+z) - \lambda(1+3 z)][\lambda +  z^2 (3+2 z) ]}.
\label{eq:Cv}
\end{eqnarray}
The schematic figure of the specific heat is shown in Fig. \ref{fig:CvsT} for $0<\lambda<1$.
Since both of the temperature and specific heat are not monotonic functions of $ z$, the figure shows a complicated structure. 
The specific heat is positive in two regions, $0< z< z_1$ and 
$ z_2 <z< z_3$.
Here $ z_1$ is the positive root of the equation
\begin{eqnarray}
 \lambda =  z^2 (3+4 z).
\end{eqnarray}
$ z_2$ is the positive root of the equation
\begin{eqnarray*}
   z^2 (3+ z) - \lambda(1+3 z) = 0,
\end{eqnarray*}
and $ {z}_3$ is defined by the point where the thermodynamic mass becomes zero, 
\begin{eqnarray}
 {z}_3 = \lambda^{1/3}. 
\end{eqnarray}
The thermodynamic mass in (\ref{thermalQ:EHM2}) has the maximum value at $ z_1$, and the specific heat vanishes.
At $ z_2$ and $ z_3$, the specific heat (\ref{eq:Cv}) diverges.
When $\lambda = 1$, however, the two roots $ z_3$ and $ z_2$ are identical, and one of the regions of $C>0$ vanishes. 
For $\lambda > 1$, the vanished region appears again, exchanging the positions of $ z_3$ and $ z_2$ each other without changing other global features. 
Thus the regions of positive specific heat for $\lambda > 1$ are $0< z< z_1$ and $z_3 < z< z_2$.

\begin{figure}
\begin{center}
  \includegraphics[width=7cm,clip]{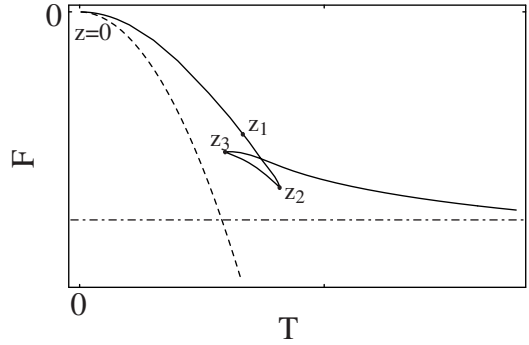}
\end{center}
\caption{ 
\label{fig:FreeEvsT}
Schematic figure of the relation between the temperature $T$ and the free energy $F$.
The BTZ black string and the black hole for $1/27 <\lambda<1$ are shown by dashed and solid lines, respectively. The free energy of the thermal AdS branes is represented by the dot-dashed line. 
In the case of $\lambda>1$, the points $ z_2$ and $z_3$ exchange their positions without changing other features (see Fig. \ref{fig:FreeEvsT2} for $\lambda >27$). 
For $0<\lambda<1/27$, the free energy of the black holes becomes less than that of the BTZ black string only around $T(z_3)$, because the left edge ($z_3$) of the triangle crosses the line of $F_{BTZ}$. 
In the limit $z\to\infty$, the free energy of the EHM II black hole asymptotes to that of the thermal AdS branes.
}
\end{figure}

When the specific heat of black holes is positive, the system can be in thermodynamically stable equilibrium. 
Then we can argue thermodynamic stability based on the free energy of the system. 
In Fig. \ref{fig:FreeEvsT}, we show a schematic figure representing the free energy of the black hole, BTZ black string and thermal AdS branes. 
The free energy of the localized black hole has a triangle region, showing that the free energy is not a monotonically decreasing function of the temperature. 
Although the triangle region is tiny in an actual plot, we have illustrated the figure stretching the part, just to show a typical behavior of the free energy. 
As the figure indicates, the free energy of the BTZ black string is less than that of the black hole for almost all temperature.
An exception happens when $\lambda < 1/27$ and $ {z} \approx  {z}_3$. 
To discuss this case, we write down the difference of the free energies at the same temperature,  
\begin{eqnarray}
F_{BH} - F_{BTZ} 
= 
\frac{2\pi}{A \kappa_4}\frac{ z^2}{
    \lambda (3 z^2 + 2 z^3 +\lambda)^2} 
\left[
  - 4 \lambda (\lambda +2 z \lambda +2  z^3 +  z^4)
  + (2\lambda +3\lambda  z + z^3)^2
\right]. 
\end{eqnarray}
When $\lambda < 1/27$, the difference $F_{BH} - F_{BTZ} \propto (-1 + 2\lambda^{1/3} + 2 \lambda^{2/3} )\lambda^{2}$ becomes negative around a temperature $T \approx T(z_3)$, where $T(z_3)$ stands for the temperature at the point $z=z_3$. 
Hence, the free energy of the black hole becomes less than that of the BTZ black string. 
Although the difference is very small, the black hole is more probable than the BTZ black string for $ T \approx T(z_3)$, 
and at this temperature the BTZ black string will decay into a black hole.
This exceptional case appears when the left edge ($z_3$) of the triangle crosses the line of $F_{BTZ}$ in Fig.\ref{fig:FreeEvsT} .  
A physical reason for this phenomenon is that in the limit $\lambda \to 0$ the BTZ black string cannot exist, but the black hole is represented by the EHM I solution.

\begin{figure}
\begin{center}
  \includegraphics[width=7cm,clip]{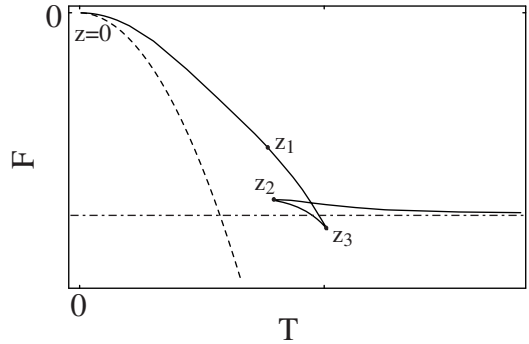}
\end{center}
\caption{ 
\label{fig:FreeEvsT2}
Schematic figure of the relation between the temperature $T$ and the free energy $F$ for $\lambda>27$. 
Only around for $T \approx T(z_3)$, a part of the free energy of the EHM II is less than that of the thermal AdS branes.
}
\end{figure}

\subsection{Comparison between the black objects and the thermal AdS branes in the canonical ensemble}

In the previous section, we have compared thermodynamic stabilities of the black objects. 
We now discuss the stabilities of the black objects with respect to the thermal branes.
Let us first consider the BTZ black string and thermal AdS branes.
The difference of free energies at the same temperature is 
\begin{eqnarray}
F_{BTZ}-F_{brane} = -\frac{2\pi}{\kappa_4 A}\left( \frac{4\pi^2}{A^2\lambda}T^2-1 \right).
\end{eqnarray}
Hence, at the temperature $T=T_c:=A\sqrt{\lambda}/2\pi$, the difference is zero. The angular periodicity at this critical temperature is $\Delta \phi_{BTZ} = 2\pi$.
For $T<T_c$, the free energy of the thermal AdS branes is less than that of the BTZ black string, and therefore, the BTZ string is less probable.
On the other hand, for $T>T_c$, the BTZ string is more probable than the thermal AdS branes.
This transition is the braneworld analogue of the three-dimensional Hawking-Page transition \cite{Hawking:1983dh, Kurita:2004yn}.

Next, we compare the EHM II solution and the thermal AdS branes.
The difference of the free energies is
\begin{eqnarray}
F_{BH}-F_{brane}=-\frac{2\pi}{A\kappa_4}\left(
\frac{4z^2(\lambda+2\lambda z+2z^3+z^4)}{(\lambda+3z^2+2z^3)^2}-1\right).
\end{eqnarray}
If $\lambda < 27$, the difference is always positive, meaning that the thermal AdS branes are more stable configuration than the EHM II black hole for any temperature. 
However, if $\lambda > 27$, an exception occurs; a local minimum of the free energy at $z_3$ becomes the global minimum (Fig. \ref{fig:FreeEvsT2}). 
Then the difference of free energy becomes negative around the point $z_3$, because $F_{brane}$ is a constant that equals the asymptotic value of $F_{BH}$ in the limit $z \to \infty$.
Therefore, for temperature $T \approx T(z_3)$, thermal AdS branes will decay into a localized black hole due to thermodynamic instability.

The last case is the EHM I black hole and the thermal flat brane.
The specific heat of the EHM I solution is always negative:
\begin{eqnarray}
C=-\frac{8\pi^2}{A^2\kappa_4}\frac{3+4z}{z(z+3)(3+2z)} < 0, \qquad {\mbox{ for}}\quad 0\le z < \infty.  
\end{eqnarray}
Thus there is no stable equilibrium as in the case of the Schwarzschild black hole.
Furthermore, the difference of free energy, $F_{BH}-F_{brane}$, is positive for any temperature.
The EHM I black hole is thermodynamically unstable in a canonical ensemble.

\subsection{Nariai instanton ($-1<\lambda <0 , k=-1$)}   

Since the cosmological constant on the brane is given by Eq. (\ref{eq:cosmological const on the brane}), the negative $\lambda$ means de Sitter brane. 
A de Sitter brane can be placed at $x=0$.
However, as we can see from Eq. (\ref{eq:tension}), we cannot place a second brane at $y=0$ as before. 
Thus we consider single de Sitter brane in this section.

From the metric on the brane, we notice that there is no black hole for $k=0, +1$, as long as we assume $m>0$ (and $A>0$ due to (\ref{eq:ell_3})).
Thus we only consider the case of $k=-1$. $F(y)$ has two negative roots for $(mA)^2 < 1/27|\lambda|$. 
The smaller root $y_0$ corresponds to a black hole horizon, and the larger one $y_1$ to a cosmological horizon. 
When $(mA)^2 = 1/27|\lambda|$, $F(y)$ has a double root, and this is an interesting extremal case where the cosmological horizon and the black hole horizon have the same size. 
Thermal equilibrium between a black hole and a heat bath is possible only in this limit. 
The limit is known as the Nariai solution, which has $\sqrt{3|\lambda|} = |y_0|$.

To take the limit of Nariai instanton, 
an expansion parameter $\epsilon$ defined by $27(mA)^2 |\lambda|=1 - 3\epsilon^2$ is conventionally introduced. 
New coordinates $(\xi,\psi)$ are $\xi = t  \epsilon \sqrt{3|\lambda| }$ and $\rho \sqrt{3|\lambda|} = 1 - \epsilon \cos \psi - \epsilon^2/6$, where $0\le \psi <\pi$ and $0\le \xi <2\pi$.
Then in the limit $\epsilon \to 0$, one finds that the metric is
\begin{eqnarray}
    ds^2
   = \frac{1}{ A^2 \bigl( x+ \sqrt{3|\lambda|} \bigr)^2} 
     \left[   \sin^2 \psi d\xi^2 + d\psi^2 + G d\phi^2 + \frac{1}{G} dx^2
     \right] . 
\end{eqnarray}
and hence on the brane (\ref{eq:metric on the x=0 EHMII}), it is the Euclidean Nariai solution, which is the product of two spheres $S^{(1)}\times S^{(2)}$.

The Euclidean action is 
\begin{eqnarray}
I 
&=&
\frac{\Delta \tau ~\Delta \phi }{\kappa_4  A^2 }
\left(
    -  \frac{ \lambda}{ y_0^2}
    +  \frac{ \lambda}{ y_1^2}
    + \frac{\lambda+1 }{(x_2-y_0)^2} - \frac{\lambda+1}{(x_2-y_1)^2}  
\right). 
\label{eq:action -deSitter}
\end{eqnarray}
By substituting the appropriate radii $y_0$ and $y_1$, the total action in the Nariai limit is 
\begin{eqnarray}
I^{N} = - \frac{\Delta \phi}{\kappa_4 A^2}
\left(  \frac{8\pi}{3} \frac{1}{\sqrt{3|\lambda| } } \right)
\left(
     1 + \frac{3\sqrt{3|\lambda|}(1-|\lambda|) }
     {   \bigl( x_2 + \sqrt{3|\lambda|} \bigr) ^3} 
\right).
\end{eqnarray}
We now compare the Euclidean action with the action for de Sitter brane without black hole. It is 
\begin{eqnarray}
I^{dS} =
- \frac{ 2\pi }{\kappa_4 A^2}
\left( \frac{2\pi }{\sqrt{|\lambda| } } \right)
\left(
     1 + \frac{ 1- |\lambda|}{(1 + \sqrt{|\lambda|})^{2}} 
\right), 
\end{eqnarray}
where we have used $y_1= - \sqrt{|\lambda|}$, $x_2=1$, $\Delta\tau = 2\pi/\sqrt{|\lambda|}$.

The difference between the two actions, 
$\Delta I= I^N -I^{dS}$, gives an evaluation of the creation rate of localized black holes during inflation \cite{Garriga:1999bq}. 
The rate is obtained by exponentiating the difference, $\Gamma \approx \exp (- \Delta I)$ \cite{Ginsparg:1983rs}. 
The difference is positive definite and it is given in the two limits by 
\begin{eqnarray}
&& \Delta I   \approx \frac{8\pi^2}{\sqrt{|\lambda|}}, 
 \quad ( |\lambda| \ll 1)
\end{eqnarray}
Thus, the creation rate becomes large when the Hubble radius ($\propto 1/\sqrt{\lambda}$) is small. 
This result is consistent with the argument given by Garriga and Sasaki \cite{Garriga:1999bq}. 
In their study, the Nariai instanton was evaluated based on the five-dimensional black string (``black cigar"), which is essentially different from that which we are interesting in.  
Although the five-dimensional black string has a curvature singularity near the tip of the ``cigar'', the singularity does not affect the evaluation of the action because the action consists of only the Ricci scalar.   
Note also that in the five-dimensional case, we can compare the instanton of the black string with the four-dimensional Nariai instanton, and we have the observation that the actions for the two instantons coincide well as long as the sizes of the instantons are sufficiently large. 
We cannot give such interesting observation to the present problem because a Schwarzschild-de Sitter solution does not exist in three-dimensions.

\section{Summary and Discussion}
\label{summary}

We have reinvestigated the thermodynamics of black objects (holes and strings) in the four-dimensional braneworld which was originally constructed and studied by Emparan et. al. 
In this paper, we have started from a different viewpoint to study the thermodynamics of the black objects.
The original investigation of the thermodynamics of the black objects was based on the assumption that there exist the first law of thermodynamics and the Bekenstein-Hawking law for the localized black hole on the brane. 
We have shown that the assumption is the case by defining the thermodynamic quantities in the Euclidean formulation, resulting in the recovery of the Bekenstein-Hawking law. 
The results for EHM II are compatible with those in Ref. \cite{Emparan:1999fd}, but those for EHM I are not the same as in Ref. \cite{Emparan:1999wa}.

Thermodynamic stabilities of black objects have been discussed in a canonical ensemble. 
We have found that the BTZ black strings are generally more stable than the localized black holes in the thermal equilibrium state with thermal radiation in the bulk. 
But an exceptional case appears when the effective cosmological constant on the brane is small enough ($\lambda < 1/27$).
In this case, the localized black hole will become more stable than the BTZ black string around the temperature $T(z_3)$.

We also found the braneworld analogue of Hawking-Page transition: the transition between the BTZ black strings and the thermal AdS two branes. 
The critical temperature is related to the AdS radius $\ell_3$ on the brane as $T_c=(2\pi \ell_3)^{-1}$.  
This relation is the same as the case of usual three-dimensional Hawking-Page transition.
Below the critical temperature $T_c$, the thermal AdS branes are more probable, while above $T_c$ the BTZ black string becomes more probable than the thermal AdS branes.  
In the comparison of the EHM II black hole and thermal AdS branes, the thermal AdS branes are generally more stable.
Only when the effective cosmological constant on the AdS brane is sufficiently large ($\lambda > 27$), will the thermal AdS branes decay into a localized black hole (EHM II) for the temperature $T \approx T(z_3)$. 
(Now $z_3$ is located in the place of $z_2$ in Fig. \ref{fig:FreeEvsT}.) 
Then the localized black hole will decay into the BTZ black string as in the above argument.

In pure 2+1 gravity, there is a three-dimensional transition between the BTZ black hole and the thermal AdS$_3$ background. In this case, conical singularities interpolate AdS$_3$ spacetime and the massless BTZ ground state.
The thermodynamic properties of such conical singularities have not been known, and we do not know how to treat such conical solutions, even though they might be important to understand the three-dimensional 
Hawking-Page transition in the vicinity of the critical temperature \cite{Kurita:2004yn}. 
In the four-dimensional braneworld, however, there are localized black holes which are counterparts to the conical singularities. 
The localized black holes ($z>z_3$) have negative specific heat and will contribute to thermodynamic instability as in the four-dimensional Hawking-Page transition \cite{Hawking:1983dh}. 
The black holes interpolate the thermal AdS branes and positive mass localized black holes. 
Hence, in this case, we can concretely see the thermodynamic behavior of the solutions interpolating the two phases.

The thermodynamic quantities for the localized black hole on the flat brane (EHM I) are different from the previously obtained ones \cite{Emparan:1999wa}. 
The thermodynamic mass that we have derived is negative contrary to the previous result. 
Since the EHM I solution can be derived in the continuous limit $\lambda \to 0$ $(z_3\to 0)$, only the negative mass sector of the EHM II solution survives in this limit (see Fig. \ref{fig:SvsM_lambda3}). 
Thus the origin of the negative mass is attributed to the negative mass sector of the black holes localized on the AdS brane.
In the minimum mass limit, the system is the four-dimensional Randall-Sundrum single brane model which is the most stable for any temperature from the thermodynamic point of view, and thus the system should be thought of as the ground state of the system.

In the usual calculation of classical actions, one has to subtract some suitable reference spacetime and obtain a thermodynamic mass relative to that of the reference spacetime.
For example, in the calculation of the action of the Schwarzschild spacetime, one usually subtracts the Minkowski background and gets finite action \cite{Gibbons:1977ue}.
For spacetimes with negative cosmological constant, one can evaluate the action without subtracting some reference spacetime: the counterterm method is useful to do it \cite{Balasubramanian:1999re,Emparan:1999pm}.
It is based on the AdS/CFT correspondence and gives finite mass even for the AdS spacetime which is often referred to as a background.
Furthermore, its mass can be interpreted as the Casimir energy of the dual CFT \cite{Balasubramanian:1999re}.
An example is well-known in AdS$_3$: 
The mass of AdS$_3$ is $-1/(8G_3)$ and this is the Casimir energy of the NS ground state of the corresponding CFT.
It is consistent with the fact that the AdS$_3$ vacuum is the NS ground state of the AdS$_3$ supergravity.
As for the black objects in the four-dimensional braneworld, we have no necessity of requiring regularization to get rid of infrared divergences in the classical action. 
As is usual, some divergences appear from each action, but the total (bare) action is still finite. 
We think that this remarkable feature would be intrinsic to the four-dimensional braneworld. 
At least, we need regularization to recover meaningful thermodynamic quantities of black strings in higher-dimensional braneworlds (Appendix \ref{appendix:BS}).

The fact that there is no need of regularization in the action, even any counterterms, implies that the value of the action and thermodynamic quantities themselves have some definite meanings.
Especially, the mass of the AdS or flat brane is the same as that of the pure AdS$_3$ spacetime.
It might be interpreted as the Casimir energy of the dual cutoff CFT.
However, we do not know any concrete description of the dual CFT for the braneworld and need further investigation to make clear this issue.
No necessity of regularization may give some insight into the corresponding CFT.

In Ref. \cite{Emparan:1999fd}, the localized black holes with angular momentum were discussed based on the AdS D-metric.
The thermodynamics of such a system will have much more complex features, and the phase diagram will become more interesting as in the case of four-dimensional Kerr-AdS black holes 
\cite{Caldarelli:1999xj}. 
We would like to defer pursuing the study to a future publication.
As we have discussed, we found the braneworld analogue of Hawking-Page transition.
A similar, but distinct ``transition" appears for the localized black holes bound on the flat brane \cite{Kudoh:2004kf};
The thermodynamic behavior changes at $T \approx (2\pi \ell)^{-1}$. 
An understanding of this phenomenon is also an interesting subject in a future study.

\begin{acknowledgments}
H.K. is supported by the JSPS. 
H.K. acknowledges the hospitality of Ishihara lab at the Osaka City University, where part of this work was completed. 
\end{acknowledgments}

\appendix

\section{$D$-dimensional black string}
\label{appendix:BS}

We are interesting in the application of the Euclidean formulation to black objects in the braneworld. 
In this Appendix we discuss a higher-dimensional black string solution as an example. The black string solution is too simple to obtain an insight into more general solution, e.g. localized black holes. 
However it is still meaningful to discuss the case for comparison with the four-dimensional black objects.

The metric of the $D$-dimensional black string in the system of two branes is 
\begin{eqnarray}
ds^2
=
\frac{1}{H^2}
\left[
f(r) d\tau^2 + f(r)^{-1} dr^2 + r^2 d\Omega_{D-3}^2+dz^2
\right] ,
\end{eqnarray}
where $D\ge 5$ and $f(r)=1- (r_h/r)^{D-4}$. 
The function $H(z)$ depends on the model of branes. 
For Minkowski branes, it is given by $H=z/\ell$ (see, e.g., Ref. \cite{Hirayama:2001bi} for other slicings).
The tensions of the two branes at $z=z_{\pm}$ are 
$\sigma_{\pm} = \pm{2(D-2)}H'(z_\pm)/\kappa_D$. 
The actions (\ref{eq:S_RS}) is 
\begin{eqnarray}
I_{EH} &=& \frac{(D-1)\beta \Omega_{D-3}
}{\kappa_D \ell^2 (D-2)}     
 \left[ r^{D-2}_\infty - r^{D-2}_h \right]
      \int^{z_-}_{z_+} dz \frac{2}{H^D},
\cr
I_{brane}+I_{GH}|_{brane} 
 &=& - \frac{ \beta \Omega_{D-3}}{ (D-2)^2 } 
       [r^{D-2}_\infty - r^{D-2}_h] 
      \sum_{k=\pm 1}  \frac{\sigma_k }{H^{D-1}(z_k)} ,
\cr
 I_{GH}|_{r=r_{\infty}} 
 &=& - \frac{ \beta \Omega_{D-3}}{ \kappa } 
      r^{D-3}_\infty
       \left[ \frac{({D-3})f}{r_\infty} + \frac{f'}{2} \right]
          \int^{z_-}_{z_+} dz \frac{2}{H^{D-2}},
\end{eqnarray}
where the temperature of the black string is given by $\beta=4\pi/f'$. 
For simplicity, we consider Minkowski branes hereafter. 
In this case we can show 
\begin{eqnarray}
    I_{EH} + I_{brane}+I_{GH}|_{brane} =0,
\end{eqnarray}
and thus only $ I_{GH}|_{r=r_{\infty}}$ gives a non-vanishing contribution to the total action. 
This is similar to the argument for the Schwarzschild black holes in asymptotically flat spacetime.  
We follow to the standard prescription of subtracting a background spacetime.
Temperature of the background spacetime is chosen to be identical to that of the black string, and it is given by $\beta_{bk}=\beta \sqrt{f}$. 
Then the finite total action is 
\begin{eqnarray}
 I &=& I_{GH} - I_{GH}^{bk}, \qquad  ({r=r_\infty})
\cr
   &=&  \frac{2 \pi \Omega_{D-3} r_h^{D-3}  }{\kappa (D-4)} 
         Z,
\end{eqnarray}
where $Z= 2 \ell^{D-2} ( z_+^{3-D} - z_-^{3-D} )/(D-3) $, 
and the second term in the first line is the action of the background RS spacetime with the temperature $\beta_{bk}$.
From Eq. (\ref{eq:M and S}), we find the energy and the entropy associated with this black string, 
\begin{eqnarray}
M &=& 
  r_h^{D-4} \frac{(D-3)\Omega_{D-3}}{2\kappa} Z,
\cr
S &=& 2\pi r_h^{D-3} \frac{\Omega_{D-3}}{\kappa} Z
  = \frac{ {\mathcal A}}{4G}.
\end{eqnarray}

\bibliographystyle{apsrev} 


\begin{thebibliography}{32}
\expandafter\ifx\csname natexlab\endcsname\relax\def\natexlab#1{#1}\fi
\expandafter\ifx\csname bibnamefont\endcsname\relax
  \def\bibnamefont#1{#1}\fi
\expandafter\ifx\csname bibfnamefont\endcsname\relax
  \def\bibfnamefont#1{#1}\fi
\expandafter\ifx\csname citenamefont\endcsname\relax
  \def\citenamefont#1{#1}\fi
\expandafter\ifx\csname url\endcsname\relax
  \def\url#1{\texttt{#1}}\fi
\expandafter\ifx\csname urlprefix\endcsname\relax\def\urlprefix{URL }\fi
\providecommand{\bibinfo}[2]{#2}
\providecommand{\eprint}[2][]{\url{#2}}

\bibitem[{\citenamefont{Myers and Perry}(1986)}]{Myers:1986un}
\bibinfo{author}{\bibfnamefont{R.~C.} \bibnamefont{Myers}} \bibnamefont{and}
  \bibinfo{author}{\bibfnamefont{M.~J.} \bibnamefont{Perry}},
  \bibinfo{journal}{Ann. Phys.} \textbf{\bibinfo{volume}{172}},
  \bibinfo{pages}{304} (\bibinfo{year}{1986}).

\bibitem[{\citenamefont{Gibbons et~al.}(2002)\citenamefont{Gibbons, Ida, and
  Shiromizu}}]{Gibbons:2002av}
\bibinfo{author}{\bibfnamefont{G.~W.} \bibnamefont{Gibbons}},
  \bibinfo{author}{\bibfnamefont{D.}~\bibnamefont{Ida}}, \bibnamefont{and}
  \bibinfo{author}{\bibfnamefont{T.}~\bibnamefont{Shiromizu}},
  \bibinfo{journal}{Phys. Rev. Lett.} \textbf{\bibinfo{volume}{89}},
  \bibinfo{pages}{041101} (\bibinfo{year}{2002}), \eprint{hep-th/0206049}.

\bibitem[{\citenamefont{Emparan and Reall}(2002)}]{Emparan:2001wn}
\bibinfo{author}{\bibfnamefont{R.}~\bibnamefont{Emparan}} \bibnamefont{and}
  \bibinfo{author}{\bibfnamefont{H.~S.} \bibnamefont{Reall}},
  \bibinfo{journal}{Phys. Rev. Lett.} \textbf{\bibinfo{volume}{88}},
  \bibinfo{pages}{101101} (\bibinfo{year}{2002}), \eprint{hep-th/0110260}.

\bibitem[{\citenamefont{Kodama}(2004)}]{Kodama:2004kz}
\bibinfo{author}{\bibfnamefont{H.}~\bibnamefont{Kodama}}
  (\bibinfo{year}{2004}), \eprint{hep-th/0403239}.

\bibitem[{\citenamefont{Ishibashi and Kodama}(2003)}]{Ishibashi:2003ap}
\bibinfo{author}{\bibfnamefont{A.}~\bibnamefont{Ishibashi}} \bibnamefont{and}
  \bibinfo{author}{\bibfnamefont{H.}~\bibnamefont{Kodama}},
  \bibinfo{journal}{Prog. Theor. Phys.} \textbf{\bibinfo{volume}{110}},
  \bibinfo{pages}{901} (\bibinfo{year}{2003}), \eprint{hep-th/0305185}.

\bibitem[{\citenamefont{Randall and
  Sundrum}(1999{\natexlab{a}})}]{Randall:1999ee}
\bibinfo{author}{\bibfnamefont{L.}~\bibnamefont{Randall}} \bibnamefont{and}
  \bibinfo{author}{\bibfnamefont{R.}~\bibnamefont{Sundrum}},
  \bibinfo{journal}{Phys. Rev. Lett.} \textbf{\bibinfo{volume}{83}},
  \bibinfo{pages}{3370} (\bibinfo{year}{1999}{\natexlab{a}}),
  \eprint{hep-ph/9905221}.

\bibitem[{\citenamefont{Randall and
  Sundrum}(1999{\natexlab{b}})}]{Randall:1999vf}
\bibinfo{author}{\bibfnamefont{L.}~\bibnamefont{Randall}} \bibnamefont{and}
  \bibinfo{author}{\bibfnamefont{R.}~\bibnamefont{Sundrum}},
  \bibinfo{journal}{Phys. Rev. Lett.} \textbf{\bibinfo{volume}{83}},
  \bibinfo{pages}{4690} (\bibinfo{year}{1999}{\natexlab{b}}),
  \eprint{hep-th/9906064}.

\bibitem[{\citenamefont{Emparan
  et~al.}(2000{\natexlab{a}})\citenamefont{Emparan, Horowitz, and
  Myers}}]{Emparan:1999wa}
\bibinfo{author}{\bibfnamefont{R.}~\bibnamefont{Emparan}},
  \bibinfo{author}{\bibfnamefont{G.~T.} \bibnamefont{Horowitz}},
  \bibnamefont{and} \bibinfo{author}{\bibfnamefont{R.~C.} \bibnamefont{Myers}},
  \bibinfo{journal}{JHEP} \textbf{\bibinfo{volume}{01}}, \bibinfo{pages}{007}
  (\bibinfo{year}{2000}{\natexlab{a}}), \eprint{hep-th/9911043}.

\bibitem[{\citenamefont{Emparan
  et~al.}(2000{\natexlab{b}})\citenamefont{Emparan, Horowitz, and
  Myers}}]{Emparan:1999fd}
\bibinfo{author}{\bibfnamefont{R.}~\bibnamefont{Emparan}},
  \bibinfo{author}{\bibfnamefont{G.~T.} \bibnamefont{Horowitz}},
  \bibnamefont{and} \bibinfo{author}{\bibfnamefont{R.~C.} \bibnamefont{Myers}},
  \bibinfo{journal}{JHEP} \textbf{\bibinfo{volume}{01}}, \bibinfo{pages}{021}
  (\bibinfo{year}{2000}{\natexlab{b}}), \eprint{hep-th/9912135}.

\bibitem[{\citenamefont{Charmousis and Gregory}(2004)}]{Charmousis:2003wm}
\bibinfo{author}{\bibfnamefont{C.}~\bibnamefont{Charmousis}} \bibnamefont{and}
  \bibinfo{author}{\bibfnamefont{R.}~\bibnamefont{Gregory}},
  \bibinfo{journal}{Class. Quant. Grav.} \textbf{\bibinfo{volume}{21}},
  \bibinfo{pages}{527} (\bibinfo{year}{2004}), \eprint{gr-qc/0306069}.

\bibitem[{\citenamefont{Kudoh}(2004{\natexlab{a}})}]{Kudoh:2004kf}
\bibinfo{author}{\bibfnamefont{H.}~\bibnamefont{Kudoh}},
  \bibinfo{journal}{Phys. Rev.} \textbf{\bibinfo{volume}{D69}},
  \bibinfo{pages}{104019} (\bibinfo{year}{2004}{\natexlab{a}}),
  \eprint{hep-th/0401229}.

\bibitem[{\citenamefont{Kudoh et~al.}(2003)\citenamefont{Kudoh, Tanaka, and
  Nakamura}}]{Kudoh:2003xz}
\bibinfo{author}{\bibfnamefont{H.}~\bibnamefont{Kudoh}},
  \bibinfo{author}{\bibfnamefont{T.}~\bibnamefont{Tanaka}}, \bibnamefont{and}
  \bibinfo{author}{\bibfnamefont{T.}~\bibnamefont{Nakamura}},
  \bibinfo{journal}{Phys. Rev.} \textbf{\bibinfo{volume}{D68}},
  \bibinfo{pages}{024035} (\bibinfo{year}{2003}), \eprint{gr-qc/0301089}.

\bibitem[{\citenamefont{Kudoh}(2004{\natexlab{b}})}]{Kudoh:2003vg}
\bibinfo{author}{\bibfnamefont{H.}~\bibnamefont{Kudoh}},
  \bibinfo{journal}{Prog. Theor. Phys.} \textbf{\bibinfo{volume}{110}},
  \bibinfo{pages}{1059} (\bibinfo{year}{2004}{\natexlab{b}}),
  \eprint{hep-th/0306067}.

\bibitem[{\citenamefont{Karasik et~al.}(2004)\citenamefont{Karasik, Sahabandu,
  Suranyi, and Wijewardhana}}]{Karasik:2004wk}
\bibinfo{author}{\bibfnamefont{D.}~\bibnamefont{Karasik}},
  \bibinfo{author}{\bibfnamefont{C.}~\bibnamefont{Sahabandu}},
  \bibinfo{author}{\bibfnamefont{P.}~\bibnamefont{Suranyi}}, \bibnamefont{and}
  \bibinfo{author}{\bibfnamefont{L.~C.~R.} \bibnamefont{Wijewardhana}}
  (\bibinfo{year}{2004}), \eprint{gr-qc/0404015}.

\bibitem[{\citenamefont{Garriga and Sasaki}(2000)}]{Garriga:1999bq}
\bibinfo{author}{\bibfnamefont{J.}~\bibnamefont{Garriga}} \bibnamefont{and}
  \bibinfo{author}{\bibfnamefont{M.}~\bibnamefont{Sasaki}},
  \bibinfo{journal}{Phys. Rev.} \textbf{\bibinfo{volume}{D62}},
  \bibinfo{pages}{043523} (\bibinfo{year}{2000}), \eprint{hep-th/9912118}.

\bibitem[{\citenamefont{Chamblin and Reall}(1999)}]{Chamblin:1999ya}
\bibinfo{author}{\bibfnamefont{H.~A.} \bibnamefont{Chamblin}} \bibnamefont{and}
  \bibinfo{author}{\bibfnamefont{H.~S.} \bibnamefont{Reall}},
  \bibinfo{journal}{Nucl. Phys.} \textbf{\bibinfo{volume}{B562}},
  \bibinfo{pages}{133} (\bibinfo{year}{1999}), \eprint{hep-th/9903225}.

\bibitem[{\citenamefont{Hawking et~al.}(2000)\citenamefont{Hawking, Hertog, and
  Reall}}]{Hawking:2000kj}
\bibinfo{author}{\bibfnamefont{S.~W.} \bibnamefont{Hawking}},
  \bibinfo{author}{\bibfnamefont{T.}~\bibnamefont{Hertog}}, \bibnamefont{and}
  \bibinfo{author}{\bibfnamefont{H.~S.} \bibnamefont{Reall}},
  \bibinfo{journal}{Phys. Rev.} \textbf{\bibinfo{volume}{D62}},
  \bibinfo{pages}{043501} (\bibinfo{year}{2000}), \eprint{hep-th/0003052}.

\bibitem[{\citenamefont{Chamblin et~al.}(2000)\citenamefont{Chamblin, Hawking,
  and Reall}}]{Chamblin:1999by}
\bibinfo{author}{\bibfnamefont{A.}~\bibnamefont{Chamblin}},
  \bibinfo{author}{\bibfnamefont{S.~W.} \bibnamefont{Hawking}},
  \bibnamefont{and} \bibinfo{author}{\bibfnamefont{H.~S.} \bibnamefont{Reall}},
  \bibinfo{journal}{Phys. Rev.} \textbf{\bibinfo{volume}{D61}},
  \bibinfo{pages}{065007} (\bibinfo{year}{2000}), \eprint{hep-th/9909205}.

\bibitem[{\citenamefont{Kodama}(2002)}]{Kodama:2002kj}
\bibinfo{author}{\bibfnamefont{H.}~\bibnamefont{Kodama}},
  \bibinfo{journal}{Prog. Theor. Phys.} \textbf{\bibinfo{volume}{108}},
  \bibinfo{pages}{253} (\bibinfo{year}{2002}), \eprint{gr-qc/0204042}.

\bibitem[{\citenamefont{Gibbons and Hawking}(1977)}]{Gibbons:1977ue}
\bibinfo{author}{\bibfnamefont{G.~W.} \bibnamefont{Gibbons}} \bibnamefont{and}
  \bibinfo{author}{\bibfnamefont{S.~W.} \bibnamefont{Hawking}},
  \bibinfo{journal}{Phys. Rev.} \textbf{\bibinfo{volume}{D15}},
  \bibinfo{pages}{2752} (\bibinfo{year}{1977}).

\bibitem[{\citenamefont{Henningson and Skenderis}(2000)}]{Henningson:1998ey}
\bibinfo{author}{\bibfnamefont{M.}~\bibnamefont{Henningson}} \bibnamefont{and}
  \bibinfo{author}{\bibfnamefont{K.}~\bibnamefont{Skenderis}},
  \bibinfo{journal}{Fortsch. Phys.} \textbf{\bibinfo{volume}{48}},
  \bibinfo{pages}{125} (\bibinfo{year}{2000}), \eprint{hep-th/9812032}.

\bibitem[{\citenamefont{Balasubramanian and
  Kraus}(1999)}]{Balasubramanian:1999re}
\bibinfo{author}{\bibfnamefont{V.}~\bibnamefont{Balasubramanian}}
  \bibnamefont{and} \bibinfo{author}{\bibfnamefont{P.}~\bibnamefont{Kraus}},
  \bibinfo{journal}{Commun. Math. Phys.} \textbf{\bibinfo{volume}{208}},
  \bibinfo{pages}{413} (\bibinfo{year}{1999}), \eprint{hep-th/9902121}.

\bibitem[{\citenamefont{Emparan et~al.}(1999)\citenamefont{Emparan, Johnson,
  and Myers}}]{Emparan:1999pm}
\bibinfo{author}{\bibfnamefont{R.}~\bibnamefont{Emparan}},
  \bibinfo{author}{\bibfnamefont{C.~V.} \bibnamefont{Johnson}},
  \bibnamefont{and} \bibinfo{author}{\bibfnamefont{R.~C.} \bibnamefont{Myers}},
  \bibinfo{journal}{Phys. Rev.} \textbf{\bibinfo{volume}{D60}},
  \bibinfo{pages}{104001} (\bibinfo{year}{1999}), \eprint{hep-th/9903238}.

\bibitem[{\citenamefont{Coussaert and Henneaux}(1994)}]{Coussaert:1994jp}
\bibinfo{author}{\bibfnamefont{O.}~\bibnamefont{Coussaert}} \bibnamefont{and}
  \bibinfo{author}{\bibfnamefont{M.}~\bibnamefont{Henneaux}},
  \bibinfo{journal}{Phys. Rev. Lett.} \textbf{\bibinfo{volume}{72}},
  \bibinfo{pages}{183} (\bibinfo{year}{1994}), \eprint{hep-th/9310194}.

\bibitem[{\citenamefont{Banados et~al.}(1992)\citenamefont{Banados, Teitelboim,
  and Zanelli}}]{Banados:1992wn}
\bibinfo{author}{\bibfnamefont{M.}~\bibnamefont{Banados}},
  \bibinfo{author}{\bibfnamefont{C.}~\bibnamefont{Teitelboim}},
  \bibnamefont{and} \bibinfo{author}{\bibfnamefont{J.}~\bibnamefont{Zanelli}},
  \bibinfo{journal}{Phys. Rev. Lett.} \textbf{\bibinfo{volume}{69}},
  \bibinfo{pages}{1849} (\bibinfo{year}{1992}), \eprint{hep-th/9204099}.

\bibitem[{\citenamefont{Emparan et~al.}(2002)\citenamefont{Emparan, Fabbri, and
  Kaloper}}]{Emparan:2002px}
\bibinfo{author}{\bibfnamefont{R.}~\bibnamefont{Emparan}},
  \bibinfo{author}{\bibfnamefont{A.}~\bibnamefont{Fabbri}}, \bibnamefont{and}
  \bibinfo{author}{\bibfnamefont{N.}~\bibnamefont{Kaloper}},
  \bibinfo{journal}{JHEP} \textbf{\bibinfo{volume}{08}}, \bibinfo{pages}{043}
  (\bibinfo{year}{2002}), \eprint{hep-th/0206155}.

\bibitem[{\citenamefont{Tanaka}(2003)}]{Tanaka:2002rb}
\bibinfo{author}{\bibfnamefont{T.}~\bibnamefont{Tanaka}},
  \bibinfo{journal}{Prog. Theor. Phys. Suppl.} \textbf{\bibinfo{volume}{148}},
  \bibinfo{pages}{307} (\bibinfo{year}{2003}), \eprint{gr-qc/0203082}.

\bibitem[{\citenamefont{Hawking and Page}(1983)}]{Hawking:1983dh}
\bibinfo{author}{\bibfnamefont{S.~W.} \bibnamefont{Hawking}} \bibnamefont{and}
  \bibinfo{author}{\bibfnamefont{D.~N.} \bibnamefont{Page}},
  \bibinfo{journal}{Commun. Math. Phys.} \textbf{\bibinfo{volume}{87}},
  \bibinfo{pages}{577} (\bibinfo{year}{1983}).

\bibitem[{\citenamefont{Kurita and Sakagami}(2004)}]{Kurita:2004yn}
\bibinfo{author}{\bibfnamefont{Y.}~\bibnamefont{Kurita}} \bibnamefont{and}
  \bibinfo{author}{\bibfnamefont{M.-a.} \bibnamefont{Sakagami}}
  (\bibinfo{year}{2004}), \eprint{hep-th/0403091}.

\bibitem[{\citenamefont{Ginsparg and Perry}(1983)}]{Ginsparg:1983rs}
\bibinfo{author}{\bibfnamefont{P.~H.} \bibnamefont{Ginsparg}} \bibnamefont{and}
  \bibinfo{author}{\bibfnamefont{M.~J.} \bibnamefont{Perry}},
  \bibinfo{journal}{Nucl. Phys.} \textbf{\bibinfo{volume}{B222}},
  \bibinfo{pages}{245} (\bibinfo{year}{1983}).

\bibitem[{\citenamefont{Caldarelli et~al.}(2000)\citenamefont{Caldarelli,
  Cognola, and Klemm}}]{Caldarelli:1999xj}
\bibinfo{author}{\bibfnamefont{M.~M.} \bibnamefont{Caldarelli}},
  \bibinfo{author}{\bibfnamefont{G.}~\bibnamefont{Cognola}}, \bibnamefont{and}
  \bibinfo{author}{\bibfnamefont{D.}~\bibnamefont{Klemm}},
  \bibinfo{journal}{Class. Quant. Grav.} \textbf{\bibinfo{volume}{17}},
  \bibinfo{pages}{399} (\bibinfo{year}{2000}), \eprint{hep-th/9908022}.

\bibitem[{\citenamefont{Hirayama and Kang}(2001)}]{Hirayama:2001bi}
\bibinfo{author}{\bibfnamefont{T.}~\bibnamefont{Hirayama}} \bibnamefont{and}
  \bibinfo{author}{\bibfnamefont{G.}~\bibnamefont{Kang}},
  \bibinfo{journal}{Phys. Rev.} \textbf{\bibinfo{volume}{D64}},
  \bibinfo{pages}{064010} (\bibinfo{year}{2001}), \eprint{hep-th/0104213}.

\end{thebibliography}

\end{document}